\documentclass[twoside,twocolumn]{article}

\usepackage{wscg}           
\RequirePackage{ifpdf}
\ifpdf
 \RequirePackage[pdftex]{graphicx}
 \RequirePackage[pdftex]{color}
\else
 \RequirePackage[dvips,draft]{graphicx}
 \RequirePackage[dvips]{color}
\fi
\usepackage{nopageno}       

\usepackage{censor} 
\usepackage{bm} 
\usepackage{physics} 
\usepackage{amssymb} 
\usepackage{mathtools}	
\usepackage{tabularx} 
\usepackage{algorithm} 
\usepackage[noend]{algpseudocode} 
\usepackage{multirow} 
\usepackage{tabularx,booktabs} 
\usepackage{todonotes}

\usepackage{xcolor}

\newcolumntype{Y}{>{\raggedleft\arraybackslash}X}

\newcommand{\schaufeli}{Excavator }
\newcommand{\schaufelii}{Sandcastle }
\newcommand{\sandburg}{Squirrel }
\newcommand{\sandhour}{Hourglass }

\StopCensoring		

\title{Interactive High-Resolution Simulation of Granular Material}

\author{
	\parbox{0.25\textwidth}{\centering
		\censor{Alexander Sommer$^{1}$}\\[1mm]
		\censor{alexander.sommer@hs-rm.de}
	}
	\hspace{0.05\textwidth}
	\parbox{0.25\textwidth}{\centering
		\censor{Ulrich Schwanecke$^{1}$}\\[1mm]
		\censor{ulrich.schwanecke@hs-rm.de}
	}
	\hspace{0.05\textwidth}
	\parbox{0.25\textwidth}{\centering
		\censor{Elmar Schoemer$^{2}$}\\[1mm]
		\censor{schoemer@uni-mainz.de}
	}
	\\[8mm]
	\parbox{\textwidth}{\centering 
		\censor{$^1$ Computer Vision and Mixed Reality Group, RheinMain University of Applied Sciences} \censor{Wiesbaden R\"usselsheim, Germany}} \\
		\censor{$^2$Institute of Computer Science, Johannes Gutenberg University Mainz, Germany}
		\vspace{5mm}
}

\usepackage{url}
\makeatletter
\def\Uslash{\mathbin{\mathchar`\/}\@ifnextchar{/}{\kern-.15em}{}}
\g@addto@macro\UrlSpecials{\do \/ {\Uslash}}
\def\Ucolon{\mathbin{\mathchar`:}\@ifnextchar{/}{\kern-.1em}{}}
\g@addto@macro\UrlSpecials{\do : {\Ucolon}}
\makeatother

\begin{document}

\twocolumn[{\csname @twocolumnfalse\endcsname

\maketitle  

\begin{abstract}
\noindent
We introduce a particle-based simulation method for granular material in interactive frame rates. We divide the simulation into two decoupled steps. In the first step, a relatively small number of particles is accurately simulated with a constraint-based method. Here, all collisions and the resulting friction between the particles are taken into account. In the second step, the small number of particles is significantly increased by an efficient sampling algorithm without creating additional artifacts. The method is particularly robust and allows relatively large time steps, which makes it well suited for real-time applications. With our method, up to 500k particles can be computed in interactive frame rates on consumer CPUs without relying on GPU support for massive parallel computing. This makes it well suited for applications where a lot of GPU power is already needed for render tasks.

\vspace{0.5em}


\vspace*{1.0\baselineskip}

\end{abstract}
}]


\section{Introduction}

\copyrightspace{80-903100-7-9}{2005}{January 31 -- February 4}

Granular materials are composed of many, small bodies that can be clearly separated from each other. The individual components of the granular material can be very different, for example grains, sand, gravel, rubble, but also beans, rice and much more.

The simulation of such materials is particularly challenging, since the macroscopic behavior of the material is determined by the microscopic interactions between the individual grains. A complete, physically correct description of all interactions is virtually impossible, so simplified models are used that accurately reflect reality to some degree. For this purpose, various methods have been developed in the field of civil engineering and later also in the field of computer graphics. While methods from the engineering field must be able to simulate acting forces as accurately as possible, the main focus of computer graphics methods is on generating visually plausible results.

A common problem with current simulation methods in computer graphics is that, despite considerable simplification, it is not possible to achieve interactive frame rates at higher particle counts that are necessary for plausible visualization. In this work we use a positionbased simulation \cite{PBDSurvey2017} with a low particle count that can be easily computed within interactive frame rates and then refine it with an efficient upsampling algorithm to obtain a large particle count. The two steps of the simulation are decoupled from each other and can be computed at different temporal resolutions. The accurate behavior of the continuum is determined by calculating the individual collisions in the first step of the simulation. The behavior of the high-resolution particles in the second step is ensured by interpolating the underlying velocity field and by partially blending in external forces.
%
%

\section{Related Work}
\label{sec:relWork}

In the field of computer science, the physical simulation of granular material plays an important role, besides physical computing, especially in computer graphics. A distinction is made between continuum methods and purely particle-based discrete methods.

Continuum methods are particularly well suited to simulate granular flow in the most time-effective manner, since there is a decoupling between grain size and resolution of the simulation. This decoupling is at the expense of finer details in areas where the motion is in free flow, for example at surfaces, free-falling grains or in the formation of spatter. Zhu and Bridson \cite{Zhu05} used a hybrid Euler-Lagrangian formulation of the Fluid-Implicit (FLIP) method, treating sand as a fluid. Narain et. al. \cite{Narain10} developed a continuum-based model that efficiently calculates internal pressures and frictional stresses with an unilateral incomprehensibility constraint. Their method also allows two-way coupling with rigid-bodies. Lenaerts and Dutr\'{e} \cite{Lenaerts09} make use of a pure Lagrangian approach by simulating granular material with the Smoothed Particle Hydrodynamics (SPH) method \cite{SPHMonaghan1992}. Aldu\'{a}n and Otaduy \cite{Alduan11} incorporated Narain et. al.'s unilateral incomprehensibility within the predictive-corrective incompressible SPH (PCISPH) method \cite{SPH09}. Klar et. al. use the Drucker-Pager plastic flow model to simulate sand in the Material Point Method (MPM) \cite{MPM95}\cite{MPM16}. Hu et. al. introduce the Moving Least Squares Material Point Method (MLS-MPM) \cite{MLSMPM18} to enable the simulation of new phenomena in MPM like two-way coupling with rigid-bodies.

Discrete methods, in contrast to continuum methods, simulate the macroscopic behavior of the material based on contacts and collisions between individual grains or particles. This enables realistic modeling of various physical phenomena. However, accurate simulation of granular media often requires a very small grain size or particle radius, and thus a large number of particles. Collision detection of a large number of particles is computationally intensive. A small particle radius usually also requires an increase of the time steps in the simulation. In practice, therefore, often unnaturally large grain sizes have to be used. The first approaches for simulating granular material with discrete methods by Cundall and Strack \cite{Cundall80} stem from Discrete Element Method (DEM) theory in molecular dynamics. Later Bell et. al. \cite{Bell05} model granular material as non-spherical particles following DEM principles with contact and shear forces for animation purposes. M\"{u}ller et. al. \cite{PBD07} developed position-based dynamics (PBD), a particle-based simulation framework that applies positional changes of particles directly to the position layer without calculating forces between individual particles. Macklin et. al. \cite{UPP14} developed a static and dynamic friction model for PBD to mimic granular material behavior in this unified framework specifically tailored for real-time applications. Fr\^{a}ncu and Moldoveanu \cite{Francu17} formulated an accurate contact and Coulomb friction model suitable for rigid and flexible bodies in PBD.    

Recent advances in the field of machine learning have also produced new AI-based approaches to predicting the behavior of granular materials. The neural networks used in these approaches are trained with classically simulated data. Again, there are approaches that use continuum methods like Coombs and Augarde \cite{Coombs20} who use MPM and Sanchez-Gonzalez et. al \cite{Sanchez20} who use SPH and MPM and approaches that use DEM like Wallin and Servin \cite{Wallin21}. Furthermore, hybrid techniques exist that combine the strengths of continuum and discrete methods, like the recently proposed method by Yue et. al. \cite{Yue18}.

In this work we focus on discrete methods. We try to overcome the weaknesses of discrete methods, namely the size of the individual particles, by splitting our simulation into two steps. First, an accurate PBD simulation that takes into account collisions between individual particles but is computed at a low resolution with large particle radii. Second, we use an efficient refinement algorithm that replaces the results of the actual simulation with a much higher number of particles with a smaller radius. The decoupling of the two simulation parts makes it possible to calculate both parts with different time steps. This way, our method is very efficient, since the upsampling in the second part can be done with much larger time steps than the contact calculation in the low resolution part. The idea of this dual partitioning is not new. It has been already applied by Aldu\'{a}n et. al. \cite{Alduan09}, who compute their low resolution (LR) guide particles using the continuum method of Bell et. al. \cite{Bell05}. They move their high resolution (HR) visualization particles using the flow of LR particles as well as external forces. Ihmsen et. al. \cite{Ihmsen2012} took up this idea. They calculated their LR particles using another continuum based method, namely the friction model in SPH developed by Aldu\'{a}n and Otaduy \cite{Alduan11}. Furthermore, they optimized the algorithm to interpolate the motion of the HR particles by superimposing external forces on the velocity field of the LR particles depending on the density of the LR particles. In contrast to the previously mentioned work, we use a discrete method with PBD. This allows real-time simulation of granular media thanks to the speed advantages of PBD over SPH. In addition, we modify the algorithm for advection of HR particles to achieve better interaction with domain boundaries and prevent particles from sticking to rigid bodies.

\begin{figure*}
	\centering
	\includegraphics[width=1.0\linewidth]{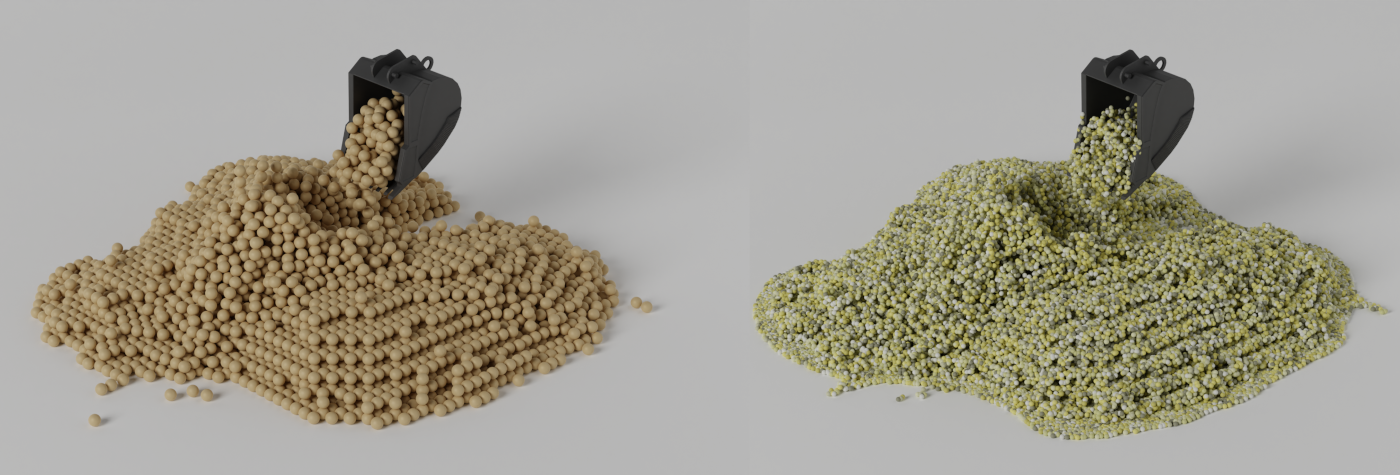}
	\caption{An excavator lifts up sand. left: LR Simulation with 6.8k Particles. right: HR Simulation with 147k.}
	\label{fg:schaufel1}
\end{figure*}

\section{LR Simulation}
\label{sec:sim}

In this section, we first describe the simulation of LR guiding particles. As mentioned before, our LR simulation is based on PBD \cite{PBD07}, respectively the modification described by Macklin et. al. \cite{UPP14} for parallel execution by constraint averaging. This allows interactive frame rates as needed in real-time applications even for larger numbers of particles.

PBD and other discrete simulation methods use particles to represent each individual discrete element of the material being simulated. All particles have the same radius $r_{LR}$. In general these particles can have arbitrary attributes. For our purposes it is sufficient to assign each particle a position $\bm{x}$, a velocity $\bm{v}$ and a mass $m$. 

In other discrete simulation methods, the particle motion is determined by time integration of all occurring internal and external forces. In PBD, however, position changes $\Delta\,\bm{x}$ due to internal forces are expressed directly by so-called constraints. The resulting velocity of individual particles is then calculated from the position difference between two time steps, i.e.
\begin{equation*}
	\Delta\,\bm{v} = \frac{\Delta\,\bm{x}}{\Delta\,t_{LR}}.
\end{equation*}
Each of the constraints used in PBD can affect $k$ particles. In general a constraint can be an equality constraint $C(\bm{x}_1, \dots, \bm{x}_k) = 0$ or an inequality constraint $C(\bm{x}_1, \dots,\bm{x}_k) \geq 0$.

For our granular LR simulation we use the friction model proposed by Macklin et. al. \cite{UPP14}. Their model consists of two different contact constraints between a collision pair $\bm{x}_i$, $\bm{x}_j$. First, interpenentrations between the two collision partners are solved by displacing the two particles along the collision vector $\bm{x}_{ij} = \bm{x}_{j} - \bm{x}_{i}$. The displacement is proportional to the mass ratio:
\begin{align}
	\Delta\,\bm{x}_i &= - \hat{m}_{ij} \left( 2r_{LR} - \abs{\bm{x}_{ij}}\right) \cdot \frac{\bm{x}_{ij}}{\abs{\bm{x}_{ij}}} \label{eq:penDisp1}\\
	\Delta\,\bm{x}_j &= \hat{m}_{ji} \left( 2r_{LR} - \abs{\bm{x}_{ij}}\right) \cdot \frac{\bm{x}_{ij}}{\abs{\bm{x}_{ij}}}\label{eq:penDisp2}
\end{align}
with 
$$
\hat{m}_{ij} = \frac{m_i^{-1}}{m_i^{-1} + m_j^{-1}}
= \frac{m_j}{m_i + m_j} 
$$
After resolving collisions the particles $i,j$ have new temporary positions $\bm{\tilde{x}}_i= \bm{x}_i +\Delta\,\bm{x}_i$, $\bm{\tilde{x}}_j=\bm{x}_j +\Delta\,\bm{x}_j$. The relative displacement between original positions and these new positions 
\begin{equation*}
	\Delta\,\bm{x}_{ij} = \left(\bm{\tilde{x}}_i - \bm{x}_i \right) - \left(\bm{\tilde{x}}_j - \bm{x}_j \right) = \Delta\,\bm{x}_i - \Delta\,\bm{x}_j
\end{equation*}
is used to calculate a frictional position delta. It is based on the tangential component
\begin{equation*}
\Delta\,\bm{x}_{ij_\perp} = \Delta\,\bm{x}_{ij} - \left( \Delta\,\bm{x}_{ij} \cdot \bm{x}_{ij} \right) \frac{\bm{x}_{ij}}{\abs{\bm{x}_{ij}}^2} 
\end{equation*}
relative to the collision vector $\bm{x}_{ij}$. With this tangential component the positional delta for particle $i,j$ is calculated as
\begin{alignat}{3}
	&\Delta\,\bm{x}_i &&= -&&\hat{m}_{ij}
	\begin{cases}
		\Delta\,\bm{x}_{ij_\perp} 	& 
		\text{if }\abs{\Delta\,\bm{x}_{ij_\perp}} < 2r_{LR} \mu_s  \\
		\Delta\,\bm{x}_{ij_\perp} \text{min}_\text{fric} &
		\text{else}
	\end{cases} 
	 \label{eq:fricDisp1}\\
	&\Delta\,\bm{x}_j &&= &&\hat{m}_{ji}
	\begin{cases}
		\Delta\,\bm{x}_{ij_\perp} 	& 
		\text{if }\abs{\Delta\,\bm{x}_{ij_\perp}} < 2r_{LR} \mu_s  \\
		\Delta\,\bm{x}_{ij_\perp} \text{min}_\text{fric} &
		\text{else}
	\end{cases}
	\label{eq:fricDisp2}
\end{alignat}
with 
$$
\text{min}_{\text{fric}} = \min{\left(\frac{2r_{LR} \cdot \mu_k }{\abs{\Delta\,\bm{x}_{ij_\perp}}},1\right)}.
$$
The parameters $\mu_s$ and $\mu_k$ are the dry friction coefficients for static and kinetic friction. For most cases, we use $\mu_s = 0.35$ and $\mu_k = 0.3$ in our simulation. For a collision pair with differing friction coefficients the cross friction coefficients
\begin{equation*}
	\mu_{ij} = \sqrt{\mu_i \mu_j}
\end{equation*}
has to be calculated, where $\mu_i$ and $\mu_j$ are the static respectively kinetic friction coefficients for particle $i$ and $j$. The static and kinetic cross friction coefficients are used in this case to calculate the positional deltas $\Delta\,\bm{x}_i$ and $\Delta\,\bm{x}_j$.

While Eq. (\ref{eq:penDisp1}) and (\ref{eq:penDisp2}) resolve only the interpenetration of the particles, Eq. (\ref{eq:fricDisp1}) and (\ref{eq:fricDisp2}) model the friction effects. In the first case of Eq. (\ref{eq:fricDisp1}) respectively (\ref{eq:fricDisp2}), static friction effects are modeled by preventing any tangential motion if the particle velocity is below a traction threshold. Kinetic friction effects are treated in the second case by limiting the positional delta depending on the penetration depth.

The friction model described previously is used to calculate the friction effects between two particles. To model the interaction between particles and rigid-bodies or domain boundaries we sample them with particles as well. This allows us to use a unified collision handling and to work with the same constraints for all types of contacts. In the \cite{Sommer21}, a method is described to sample arbitrary closed volumes of 3D triangle meshes with particles for the needs of particle-based simulation. For stationary objects, the particles are assigned an infinitely large mass or an inverted mass of $m^{-1} = 0$, respectively. In this case it is sufficient to only sample the surface of the object. A method to sample arbitrary surfaces of 3D triangle meshes based on the uniform sampling algorithm of Bowers et. al. \cite{Bowers10} can also be found in \cite{Sommer21}. For moveable rigid-bodies the individual particles are first treated as if they were unconnected. Then, a shape-matching constraint \cite{ShapeMatching05} is applied to the particles of the rigid body to find the particle configuration corresponding to a transformed resting state. The position delta from this constraint is given by
\begin{equation*}
	\Delta\,\bm{x}_i = \left( \bm{R} \bm{x}_{o_i} + \bm{c} \right) - \bm{\tilde{x}}_i
\end{equation*}
where $\bm{c}$ corresponds to the center of gravity of the new deformed particles of the rigid body. $\bm{x}_{o_i}$ corresponds to the offset of the i-th particle from the center of gravity of the undeformed particle positions. $\bm{R}$ is a rotation matrix which is given by the polar decomposition \cite{PolarDecomposition} of the covariance matrix
\begin{equation*}
	\bm{C} = \sum_i^{n} \left( \bm{\tilde{x}}_i - \bm{c} \right) \cdot \bm{x}_{o_i}^T
\end{equation*}
of the deformed shape. This enables a two-way coupling between the granular medium and rigid-bodies.

All of these previously mentioned constraints, namely contact, friction and shape-matching constraints, can be calculated in parallel. A particle $i$ can receive a position delta $\Delta\,\bm{x}_{i_k}$ from several constraints $k$. Therefore, after all constraints have been calculated, the sum of all positions delta belonging to a particle $i$ is divided by $n_i$ the number of constraints involved:
\begin{equation*}
	\Delta\,\bm{x}_i = \frac{1}{n_i} \sum_k^{n_i} \Delta\,\bm{x}_{i_k}.
\end{equation*}
The procedure of constraint solving is repeated a couple of times until a solution is found that satisfies all the constraints. For our calculations between 3 and 5 iterations were sufficient. However, it can happen that no convergence was achieved and particles are in an invalid position at the start of a simulation step. This can lead to particles experiencing an unwanted acceleration in the next time step due to the next constrain to solve. This effect is especially strong the smaller the time step is. To avoid this, 1-2 stabilization iterations of the pure contact constraints, Eq. (\ref{eq:penDisp1}) and (\ref{eq:penDisp2}), are executed per time step. The resulting position deltas are applied to the temporary positions as well as to the regular positions. This prevents unnatural kinetic energy from being added to the system due to these irregularities when calculating new velocities. The complete algorithm for performing one time-step is shown in Algorithm \ref{alg:lrsim}.

To achieve a more stable piling of particles, e.g. in sand piles, and to prevent dissolution, a further improvement is made. As described in \cite{UPP14} a scaled mass 
\begin{equation*}
	m^*_i = m_i \cdot e^{-h(\bm{x}_i)}
\end{equation*}
is assigned to each particle $i$ based on the relative height $h(\bm{x}_i)$ with respect to the ground plane. These scaled masses $m^*$ are used to solve the contact and friction constraints. Since higher particles exert less pressure on the levels below, the system converges faster and the piles are more stable.

\begin{algorithm}
	\centering
	\begin{algorithmic}[1]
		\ForAll {particles $i$}
		\State $\bm{v}_i \leftarrow  \bm{v}_i + \Delta\,t_{LR} \bm{g}$ \Comment{$\abs{\bm{g}} = 9.81 \frac{m}{s^2}$}
		\State $\bm{\tilde{x}}_i \leftarrow  \bm{x}_i + \Delta\,t_{LR} \bm{v}_i$
		\State $m^*_i \leftarrow m_i e^{-h(\bm{x}_i)}$
		\EndFor
		\State find$\_$neighboring$\_$particles()
		\State generate$\_$stabilization$\_$constraints()
		\For {stabilization iterations $it = 1,2$}
			\State $\Delta\,\bm{x} \leftarrow 0$, $n \leftarrow 0$ \Comment{$\forall$ particle $i$}
			\State $\Delta\,\bm{x}, n$ $\leftarrow$ solve$\_$stabilization$\_$constraints()
			\ForAll {particles $i$}
				\State $\bm{\tilde{x}}_i \leftarrow  \bm{\tilde{x}}_i + \Delta\,\bm{x}_i / n_i$
				\State $\bm{x}_i \leftarrow  \bm{x}_i + \Delta\,\bm{x}_i / n_i$
			\EndFor
		\EndFor
		\State generate$\_$constraints()
		\For {solving iterations $it = 1,\dots,5$}
			\State $\Delta\,\bm{x} \leftarrow 0$, $n \leftarrow 0$ \Comment{$\forall$ particle $i$}
			\State $\Delta\,\bm{x}, n$ $\leftarrow$ solve$\_$constraints()
			\State $\bm{\tilde{x}}_i \leftarrow  \bm{\tilde{x}}_i + \Delta\,\bm{x}_i / n_i$
		\EndFor
		\ForAll {particles $i$}
		\State $\bm{v}_i \leftarrow  \left(\bm{\tilde{x}}_i - \bm{x}_i \right) / \Delta\,t_{LR}$
		\State $\bm{x}_i \leftarrow  \bm{\tilde{x}}_i$
		\EndFor
	\end{algorithmic}
	\caption{One simulation time-step}
	\label{alg:lrsim}
\end{algorithm}

\section{HR Upsampling}
\label{sec:upsampling}

In the previous section we described the simulation of LR guide particles taking into account all collisions between particles. We now describe how a finer resolution simulation result can be generated on the basis of the LR particles with the help of an upsampling. Figure \ref{fg:schaufel1} shows a comparison between the LR simulation on the left and the result of the HR upsampling on the right. This is done without having to perform complex collision queries between HR particles. Thus this upsampling is extremely effective and the size of the LR time step is decoupled from the size of the HR particles. Each LR Particle can be seen as a representation of several HR Particles.
\begin{figure}
	\centering
	\includegraphics[width=1.0\linewidth]{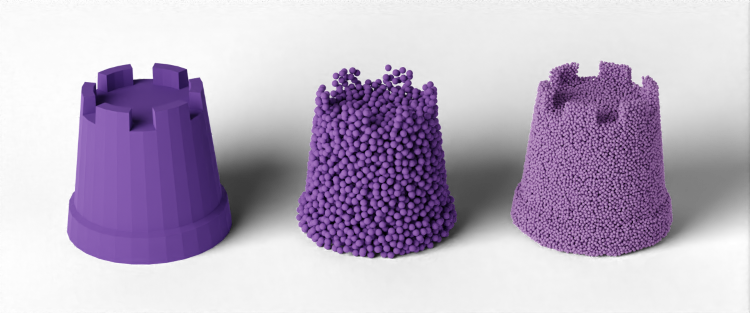}
	\caption{Exemplary illustration of the sampling process. left: Inital sandcastle mesh. middle: LR sampling. right: HR sampling }
	\label{fg:sampling}
\end{figure}

It is especially important to create a good initial sampling to avoid artifacts. For example, a uniform sampling can lead to repetitive patterns during the simulation. Furthermore, if the HR particles are placed symmetrically around the LR particles, aliasing or staircase patterns can occur already in the initial state. Therefore, we use the randomized volume sampling algorithm described in \cite{Sommer21} for both the LR and the HR particles. We use a triangle mesh as a hull in which the particles should be located. First the bounding box delimiting the mesh is divided into grid cells with a side length of $\frac{2 r}{\sqrt{3}}$, with $r$ being the radius of the LR particles $r_{LR}$ respectively HR particles $r_{HR}$. Then, a large number of randomly but uniformly distributed candidate positions are generated within the mesh and assigned to the respective grid cells. After that, for each grid cell within the mesh, an attempt is made to select a candidate position that does not overlap with already sampled positions. In this fashion the whole volume of the mesh is sampled. For a more detailed description we refer the reader to \cite{Sommer21}. This algorithm is executed for both LR particles with radius $r_{LR}$ and HR particles with radius $r_{HR}$ before starting the simulation. Figure \ref{fg:sampling} shows an example of the sampling process. The initial mesh is displayed on the left, while in the middle the sampling for the LR simulation and on the right the corresponding sampling for the HR upsampling is shown.

\begin{figure}
	\centering
	\includegraphics[width=1.0\linewidth]{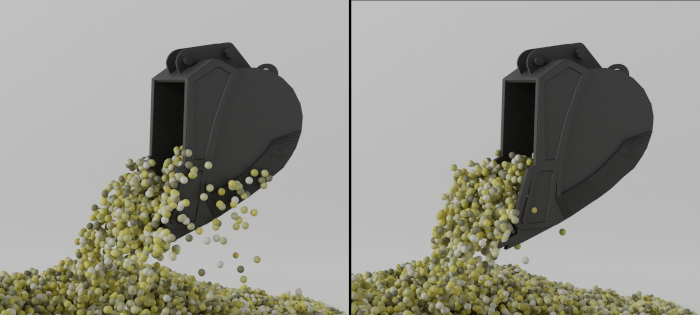}
	\caption{left: Ihmsen et. al.'s original upsampling with sticking artifacts. right: HR Simulation with our modification. }
	\label{fg:compareIhmsen}
\end{figure}

Once the simulation is started, the LR particles are set in motion by the LR simulation described in the previous section. The HR particles trace the movement of these LR guide particles. HR particles should follow the flow of the LR simulation but still move individually to avoid the formation of clumps. For this we use the advection method described by Ihmsen et. al. \cite{Ihmsen2012}. For this, gravity is smoothly faded in as an external force depending on how densely a HR particle is surrounded by LR particles. If a HR particle is in the vicinity of many LR particles, the velocity field generated by them dominates. For this, distance-based weights $w$ between HR particles $i$ and LR particles $j$ are calculated as
\begin{equation*}
	w_{ij}= \max \left( 0, \left(1 - \frac{\abs{\bm{x}_{ij}}^2}{9 \cdot r_{LR}^2}\right)^3\right).
\end{equation*}
These weights are used to determine the average velocity $\bm{v}_i$ at a HR particle $i$ with
\begin{equation*}
	\bm{\tilde{v}}_i = \frac{1}{\sum_j w_{ij}} \sum_j w_{ij} \bm{v}_j .
\end{equation*}
It shall be noted that LR particles $j$ can be granular material as well as the particle representation of rigid-bodies or boundaries. The blending in of external forces is controlled by a parameter
\begin{equation*}
	\alpha_i = \begin{cases}
	1 - \max w_{ij} &
	\text{if }\max w_{ij} \leq c_1 \text{ or } \frac{\max w_{ij}}{\sum_j w_{ij}} \geq c_2\\
	0 &\text{else}
	\end{cases}
\end{equation*}
which differs from zero only in sparse regions. The constant $c_1=\frac{512}{729}$ is the distance-based weight for a distance equal to the LR particle radius $r_{LR}$ and $c_2=0.6$ is an empirically tested value. With this blending parameter the resulting velocity for a HR particle $i$ is calculated as
\begin{equation*}
	\bm{v}_i^{t + 1} = (1 - \alpha_i) \bm{\tilde{v}}_i^{t+1} + \alpha_i \left( \bm{v}_i^t + \Delta\,t_{HR} \bm{g}\right)
\end{equation*}
where $\bm{g}$ is the acceleration due to gravity and $\Delta\,t_{HR}$ denotes the time-step size from time $t$ to time $t+1$. The new position $\bm{x}_i$ for the $i$-th HR particle is then trivially obtained by time integration through
\begin{equation*}
	\bm{x}_i^{t+1} = \bm{x}_i^t + \Delta\,t_{HR} \bm{v}_i^{t + 1}.
\end{equation*}
In contrast to Ihmsen et. al. we ignore the LR particles of rigid bodies if there are no LR granular particles in the vicinity. This prevents HR particles from sticking to rigid bodies that are moved externally (see Figure \ref{fg:compareIhmsen}).

\section{Results}
\label{sec:results}
In contrast to previous works, our work focuses on runtimes in the range of interactive frame rates. Therefore, in this section we present some examples of our approach to illustrate the different possible applications. 

\subsection{Implementation}
We have implemented our simulation framework in C++ 14. For more complex mathematical operations and mathematical data structures like vectors and matrices we use the Eigen3 \cite{Eigen3} library. We parallelized our algorithms on the CPU using OpenMP \cite{OpenMP}. For an efficient neighborhood search we use the CompactNSearch library based on the Compact Hashing approach by Ihmsen et al. \cite{Ihmsen11}. Furthermore, we use the \textsc{Leaven} \cite{Sommer21} library for sampling volumes and surfaces. All renderings in this work were created with Cycles in Blender \cite{Blender}.

\subsection{Experiments \& Performance}
In this section we would like to underline and evaluate the usefulness of our method by suitable experiments (scenes). For this purpose, we have selected four different test scenarios. For all scenarios we use a default time step $\Delta t_{LR} = 0.005 \text{s}$ for the LR simulation, which is further constrained by the CFL condition if necessary. The time step for upsampling is $\Delta t_{HR} =  0.0167 \text{s}$. The density of the granular medium is $\rho = 1600 \frac{\text{kg}}{\text{m}^3}$ and the kinetic and static friction coefficients are $\mu_k = 0.3$ and $\mu_s = 0.35$, respectively. In general, we use LR simulation radii $r_{LR}$ between 0.005m and 0.05m. For the upsampling we use a HR particle radius between $r_{HR} = 0.2 \cdot r_{LR}$ and $r_{HR} = 0.4 \cdot r_{LR}$. This leads to an upscaling factor between 15 and 125.
\begin{table}
	\begin{tabularx}{\columnwidth}{|X|c c|c c|}
		\hline
		& \multicolumn{2}{c|}{Particles} & \multicolumn{2}{c|}{Time per Frame}  \\
		& LR & HR & LR & HR \\
		\hline
		\schaufelii & 2.4k & 90k & 9.75ms & 7.31ms \\
		\hline
		\schaufeli & 6.8k & 147k & 21.4ms & 11.2ms \\
		\hline
		\sandhour & 10k & 460k & 47.4ms & 48.5ms \\
		\hline
		\sandburg & 10k & 207k & 33.0ms & 15.8ms \\
		\hline
	\end{tabularx}
	\caption{Timing results for four different scenes}
	\label{tbl:timing}
\end{table}

\begin{figure}
	\centering
	\includegraphics[width=1.0\linewidth]{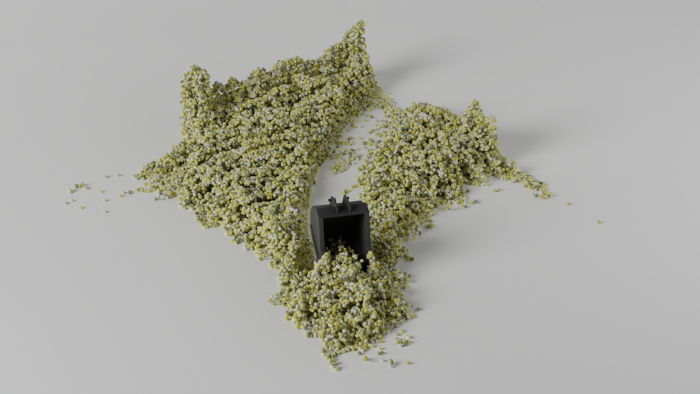}
	\caption{An excavator is moved by user input through a sand castle}
	\label{fg:schaufel2}
\end{figure}

Scene \schaufelii (see Figure \ref{fg:schaufel2}) consists of a excavator controllable by user interaction and a sand castle with a relatively small number of particles. The LR particle radius is 0.03m and the upsample radius is 0.01m, resulting in 2.4k LR particles to 90k HR particles. This scene serves to demonstrate the real-time capability of our method, in which relatively few elaborately simulated particles can be used to obtain visually pleasing results by upsampling.

Scene \schaufeli, shown in Figure \ref{fg:schaufel1}, also contains an excavator. In this example, a radius of 0.02m generates almost three times as many LR particles (6.8k) as in \schaufelii. The upsampling radius 0.008m generates 147k HR particles. This scene is intended to illustrate that our method can be used to simulate even complicated interactions such as the lifting and lowering of granular material with an excavator.

The \sandhour example should on one hand clarify the behavior and runtime of larger particle numbers and on the other hand, more importantly, give an impression of the stable piling of our method. For this purpose 10k LR particles in the hourglass are upsampled to 460k HR particles. Radii of 0.028m and 0.008m are used.

Scene \sandburg shows the two-way coupling between rigid bodies and the granular material. For this purpose, a squirrel is hurled into the sandcastle with an initial velocity. The squirrel is represented by 1.4k particles with a density of $\rho = 5000 \frac{\text{kg}}{\text{m}^3}$, which are simulated in the LR simulation. Afterwards, the original squirrel shape is reconstructed by shape matching as described in Section \ref{sec:sim}, which accomplishes the two-way coupling. All LR simulation particles have a radius of 0.02m and the upsampling radius of the granular particles is 0.0075m. In addition to the 1.4k particles of the squirrel, there are 10k LR granular particles in the simulation, which are upsampled to 207k.

All scenes were calculated on an Intel Core i9-9980HK with 8 cores on the CPU alone. Table \ref{tbl:timing} shows the per frame timing results for the four different scenes. For the timing of the LR simulation, the run-times of as many simulation steps as necessary to calculate a time difference $t = \Delta t_{HR} =  0.0167 \text{s}$ were combined. It shows that even for relatively high particle counts of 500k a computation within interactive frame rates is possible with our method on consumer hardware without exploiting massive parallelism on GPUs. Especially where the resources of the GPU are already needed for e.g. computationally intensive physically based rendering for realistic visualization, our pure CPU based simulation can show its advantages.

\begin{figure}
	\centering
	\includegraphics[width=1.0\linewidth]{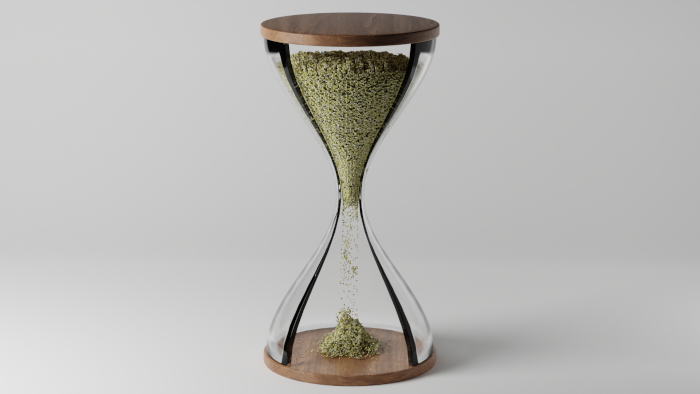}
	\caption{An hourglass with sand containing 460k particles.}
	\label{fg:hourglass}
\end{figure}

\begin{figure}
	\centering
	\includegraphics[width=1.0\linewidth]{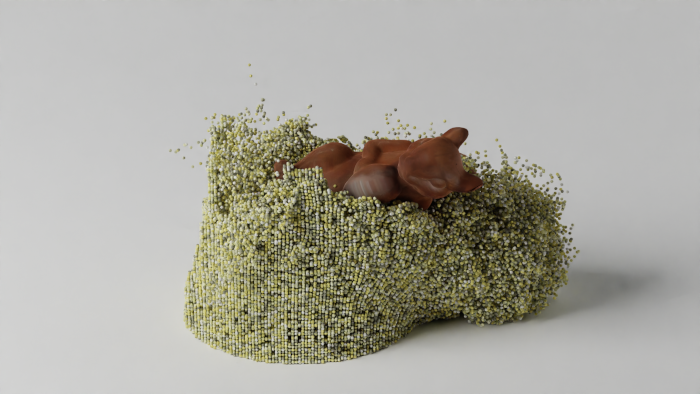}
	\caption{A squirrel hurled into a sand castle.}
	\label{fg:squirrel}
\end{figure}
\section{Conclusion \& Future Work}
\label{sec:conclusion}
With this work we have shown that with the help of an upsampling algorithm it is possible to scale up relatively small numbers of particles to produce visually vivid results within interactive frame rates for the simulation of granular material. It was shown that the upsampling method described by Ihmsen et. al. \cite{Ihmsen2012} for SPH simulations, which are far away from real-time computation times, is very well suited for application in interactive PBD simulations. We believe that this is advantageous for future real-time applications in interaction with granular material. 

\section{Limitations \& Future Work}
On one hand, in position-based simulations friction is dependent of the number of iterations, which has a negative impact on the correctness of the friction effects in the LR simulation part. The modeled static and kinetic friction effects are only approximations. On the other hand, static friction cannot be modeled correctly with HR upsampling, which prevents stable piling in some situations.

Furthermore, currently only one constant radius for all particles in the LR simulation and another constant radius for all particles in the HR upsampling is possible.

In the future, we plan to take advantage of the high parallizability of the PBD variant described by Macklin et. al. \cite{UPP14} and the trivial parallelizability of the upsampling algorithm to create a GPU-based version of our simulation that can simulate even larger numbers of particles in real time.

\section*{ACKNOWLEDGMENTS}
\censor{This Project is supported by the Federal Ministry for Economic Affairs and Climate Action (BMWK) on the basis of a decision by the German Bundestag.}


\bibliographystyle{plain}

\footnotesize
\bibliography{references}

\end{document}